# Governing Mental-State Inference: Source-Neutral Regulatory Triggers and Tiered Obligations

**Shinnosuke Horiuchi**
Department of Education, College of Arts, Rikkyo University, Tokyo, Japan | s-horiuchi@rikkyo.ac.jp

## Abstract

Two systems can supply the same person-linked attribution to the same institutional decision maker yet fall into different legal categories: one uses neural signals, the other text or behaviour. A source-bound rule therefore permits circumvention, while an all-purpose category of “mental data” risks treating fallible outputs as facts about the mind. This article reads the 2025 United Nations Educational, Scientific and Cultural Organization (UNESCO) Recommendation on the Ethics of Neurotechnology as non-binding guidance and develops a source-neutral trigger for technologically mediated, person-linked mental-state attribution. Through selective critical synthesis, conceptual engineering, functional legal comparison, and matched counterfactual cases, it separates elicitation, attribution, and use as cumulative objects of regulation. The analysis also distinguishes two harm pathways from two independently assessed duty series. Seven ordered questions and two escalation predicates assign permitted practices to three duty tiers. Presumptive prohibition is reserved for materially autonomy-affecting non-consensual closed-loop intervention and for specified combinations of covert or coercive inference, core attributes, consequential decisions, and manipulation. Common entry does not erase source-related aggravators: invasiveness, embodiment, and closed-loop capacity can add duties or set a higher minimum, while consequential non-neural inference can reach the same tier. The resulting scheme is a classification device, not an empirically validated regulatory outcome.

**Keywords:** mental-state inference; mental privacy; neurotechnology; large language models; UNESCO; risk-based regulation

## 1. Functionally Comparable Profiles, Fragmented Protection

An employer receives a score classifying an applicant as emotionally unstable. In one version of the case, the score is produced from neural signals recorded through a wearable device. In another, it is inferred from the applicant's writing and prior online posts. Assume that the attributed property, output format, accuracy range, recipient, employment decision, and retention period are held constant. The first version may enter a statute or policy framed around neural data. The second may fall instead under general data protection, an artificial intelligence (AI) rule limited to particular inputs or uses, or no special category at all. The applicant nevertheless confronts the same institutional object: a person-linked attribution presented as evidence about a mental characteristic and used by a gatekeeper.

The comparison does not presuppose that either system discovers the same inner fact. Facial movements sometimes covary with emotion categories, but the reviewed evidence does not support treating particular configurations as reliable, specific, and context-independent diagnostic displays [16, pp. 1–2]. Reverse inference from neuroimaging depends on selectivity and prior probabilities [20, pp. 59–61, Table 1]. In one static-text study, GPT-3.5 and GPT-4 estimates of Big Five traits from archived Facebook posts correlated modestly with self-reports, with a mean correlation of .29 and trait-level correlations from .22 to .33 [19, Results, Figs. 1–2]. That study did not test live conversation or downstream gatekeeping. Han and Chen instead reconstruct existing brain-computer interface (BCI) studies to argue that mind-readability is co-produced by task design, calibration, embodied engagement, cooperation, and resistance [31, pp. 6–11]. The object that reaches an institutional decision is thus not a mental state itself, but a socially operative attribution whose epistemic status and practical effects require scrutiny.

Source-specific protection remains justified for some purposes. Neural acquisition may be invasive, closely coupled to the body, difficult to avoid, or integrated into a feedback system capable of intervening in real time. These features can intensify mental-privacy and mental-integrity risks [11, 12, 21, 24]. They do not, however, furnish a complete general boundary. Balliu and colleagues predicted within-person depression-severity trajectories from passive smartphone features in a longitudinal cohort [18, pp. 1–2, 6–7], while Peters and Matz obtained modest trait-score associations from archived Facebook histories [19, Results, Fig. 2 and Limitations]. Martinez-Martin and colleagues offer a different kind of evidence: their Perspective explains why ostensibly content-free digital-phenotyping inputs may generate sensitive inferences and corresponding ethical questions [17, pp. 1–3]. None of these studies tests institutional gatekeeping or personalised manipulation. Their findings nevertheless expose the omission created when technically produced attributions count only if they originate in neural signals. The opposite mistake is to label every probabilistic output "mental data", thereby legitimating weak science, overregulating ordinary communication, and converting contestable interpretations into administratively certified facts.

Ontological identity between neural and non-neural inputs is beside the point; the inputs plainly differ. The regulatory question is: **How should regulation identify a common entry point for technologically mediated, person-linked mental-state attribution while differentiating duties according to epistemic weakness, intrusion, institutional power, and the consequences of use?**

To answer it, this article locates the protected regulatory object in a technologically generated or materially shaped person-linked attribution, not in a metaphysically uniform substance called mental data. Input and elicitation conditions, attribution outputs, and deployment or institutional use remain distinct, with duties attaching cumulatively as a system crosses each stage. Seven ordered Yes/No questions, supplemented by core-attribute and personalised-manipulation predicates, allocate permitted practices across three duty tiers and identify practices subject to presumptive prohibition. Epistemic duties and intrusion-and-power duties are assessed separately but may apply together: an inaccurate inference and an accurate inference exploited coercively pose different problems, and either can engage both sets of duties. The contribution lies in how these elements are combined and ordered as a decision architecture, not in a claim that any one element is unprecedented.

The claim is deliberately limited. This article neither creates a new human right nor offers empirically validated thresholds, a ranking of national legal orders, or a complete implementation plan for any jurisdiction. Privacy, data protection, mental integrity, freedom of thought, equality, and due process all appear in the relevant literature, but their scope and legal form remain contested [12–15, 23–27]. The authority for implementation must therefore be identified jurisdiction by jurisdiction. What the model supplies is a classification and allocation device that existing or newly specified authorities could use. Section 2 sets out the method; Sections 3 and 4 reconstruct the source-bound debate and UNESCO's implementation problem; Sections 5 and 6 develop the regulatory objects and duties; and Sections 7 and 8 test the proposal against existing legal forms and its strongest objections.

## 2. Analytical Approach

The method combines problem reconstruction with institutional testing. Selective critical synthesis identifies positions that a workable governance model must preserve or answer, while conceptual engineering clarifies the object to which regulation should attach. The sources were drawn from neuroethics, neurorights, mental privacy, brain-data governance, and mental-state-inference research because they address the protected interest, the epistemic status of inference, the regulatory object, or the institutional form of protection. The review is selective rather than systematic and does not estimate how widely any position is held.

Conceptual engineering is required because terms such as *brain data*, *neural data*, *mental data*, *mind reading*, and *mental privacy* do not merely describe stable technical kinds. They allocate visibility, duties, and standing. The proposed trigger must catch functionally comparable person-linked attributions without implying direct access to a pre-existing inner state. It must also preserve differences in acquisition, calibration, uncertainty, reversibility, embodiment, and use. "Mental-state attribution" names the representational output, "mental-state inference" the process, and "regulatory trigger" the common legal entry point.

Institutional testing then combines functional legal comparison with matched counterfactual cases. The comparison concerns five regulatory types rather than supposedly commensurable legal systems: a cross-source international recommendation, a general data-protection regime, a use- and input-specific AI regime, neural-data statutes, and constitutional protection focused on brain activity. This is not a representative sample of global law. The instruments were chosen because they expose different points at which a system may enter or escape protection. Throughout the comparison, current binding law, authoritative non-binding guidance, interpretive reconstruction, and the author's proposal remain distinct.

Pair A holds constant the person, attributed property, output form, accuracy range, user, institutional decision, and retention period while changing only whether the input is neural or non-neural. Pair B is a three-variant aggravation test. B1 adds only invasive acquisition, B2 only close embodied integration, and B3 only closed-loop capacity to the neural version. Each variant retains Pair A's output and institutional purpose. These are normative counterfactuals, not empirical measurements of a single true mental content. Pair A tests common entry. The B variants illustrate the model's proposed treatment of source-related aggravators without attributing an observed causal effect to any one feature.

The proposed tiers and escalation rules are not scientifically calibrated thresholds. The underlying materials perform different roles. Empirical studies document specific non-neural inference practices [18, 19]; review and methodological analyses identify epistemic limitations [16, 20]; Martinez-Martin and colleagues map ethical questions raised by digital phenotyping [17]; and Han and Chen reconstruct BCI-mediated decoding conceptually [31]. The official instruments establish current legal categories, obligations, and policy recommendations [1, 4–8], while normative and governance studies identify protected interests, institutional alternatives, and objections [9–15, 21–32]. The decision architecture that connects these materials is the article's institutional design and remains open to criticism.

In revising the manuscript, the author used AI tools to suggest alternative English formulations and to assist in locating relevant passages and checking citations within a pre-existing collection of research materials assembled and categorized by the author. The assistance with English formulation included suggestions for language and copy editing of author-written text, with the aim of preserving the author's intended meaning while improving clarity and readability. The author reviewed all suggested wording, made all final decisions concerning the text, framework, argument, and the selection and interpretation of sources, verified factual and legal claims against the cited scholarly, legislative, judicial, or official materials, and takes full responsibility for the final manuscript.

## 3. From Neuroexceptionalism to Functional Protection

Neurorights scholarship begins from a persuasive concern: techniques that record, decode, or intervene in neural activity may expose or alter aspects of mental life in ways that existing legal categories fail to capture. Proposals for mental privacy, mental integrity, cognitive liberty, and psychological continuity have sought to specify that concern [12, 13]. Responsible-neurotechnology frameworks likewise emphasize safeguards across design, research, deployment, and governance [2, 11, 21]. These approaches preserve the embodied character of neural intervention and the possibility that a device may become integrated with action, communication, or self-understanding.

The disagreement concerns what legal consequences should follow from those risks. Some proposals create new rights or neural-data categories. Others seek to interpret or reorganise existing human-rights and data-protection rules [13, 14, 23], while a further line of argument centres the capacity to infer, manipulate, or govern mental life rather than the formal source of the data [9, 27, 32]. Invoking technological novelty does not resolve the choice among them. The institutional question is which boundary protects the relevant interest without becoming unadministrable.

A neural-source boundary offers genuine administrative clarity. It identifies a technically recognisable acquisition domain, keeps embodiment and potential invasiveness in view, and resists extending mental privacy to every observation from which one person might form an impression of another. California's amended definition, for example, expressly distinguishes neural data generated by measuring nervous-system activity from information inferred from non-neural sources [6]. Colorado likewise anchors neural data in device-processable measurements of central or peripheral nervous-system activity [7]. Chile's constitutional text directs legislation governing scientific and technological use in persons especially to safeguard brain activity and information originating from it [8]. Regulators and regulated actors can therefore locate the protected object near its source.

The same clarity produces an avoidance route. If a later institution relied on one of the person-linked non-neural attributions described above, a source-only boundary could subject comparable gatekeeping to weaker scrutiny. This is a regulatory possibility, not an outcome tested by the cited studies. Brown distinguishes natural, digital, and neurotechnological forms of mind reading while resisting protection limited to neurotechnology [9]. Input source alone does not determine whether a mental-privacy interest is engaged.

An expert paper prepared for the Convention 108 Committee likewise discusses a functional category spanning neural and non-neural routes to mental or cognitive information [3]. It is not a treaty text, a Committee recommendation, or an authoritative Council of Europe interpretation; its cover states that the authors' views do not necessarily reflect official policy. Its relevance here is conceptual, not legal.

Broadening the category to mental data would close that gap only by creating another. A model output is not a detachable sample of the mind. Research on facial emotion inference shows how context, culture, and within-category variation complicate the move from observable movement to a discrete emotional state [16, pp. 1–2]. Poldrack makes an analogous point about neuroimaging: activation in a region does not deductively establish a single cognitive process, although appropriately specified probabilities may still be informative [20, pp. 59–61]. Han and Chen press the objection further by presenting neural decoding as an interactive accomplishment shaped by task construction, stimuli, individual calibration, embodied history, and the participant's cooperation or resistance [31, pp. 6–11]. Treating the output as self-authenticating mental content would hide the conditions that produced it.

Protection can instead attach to a socially operative attribution without certifying it as true. Such an attribution may affect employment, health care, policing, insurance, or personalised persuasion even when it is uncertain, disputed, or wrong. Its regulatory relevance comes from technological production, person-linkage, an assigned mental meaning, and practical use. The claim concerns institutional operation, not the ontology of mind.

Functional protection must nonetheless retain source-related differences. López-Silva and colleagues argue that neurotechnological applications may create distinctive mental-privacy risks, while cautioning that distinctiveness does not by itself settle whether a new right or instrument is required [24]. The appropriate inference is conditional. Neural origin is neither an automatic sufficient reason for the highest protection nor an irrelevant detail. It can aggravate risk where acquisition is invasive, where processing is closely embodied, where continuous monitoring is difficult to refuse, or where inference feeds directly into intervention. Pair B is designed to make these aggravators visible after common coverage has been established.

This position also limits the legal form of the proposal. A source-neutral trigger need not become a freestanding universal right or a single new statute. It may operate as a classification rule within data protection, sectoral AI law, employment law, medical-device regulation, consumer protection, equality law, or public-law due process. Hohfeldian analysis warns against treating every protected interest as the same kind of claim-right [23]. Structural accounts add that individual consent and complaint may be insufficient when platforms or public authorities organise the conditions under which cognition is inferred and acted upon [32]. A functional trigger must therefore be portable across legal forms while remaining precise enough to allocate identifiable duties.

## 4. UNESCO as a Normative Bridge

The 2025 UNESCO Recommendation on the Ethics of Neurotechnology does not supply a ready-made regulatory scheme. It is an authoritative but non-binding standard-setting instrument that can guide national law and practice; its substantive provisions are not directly enforceable global rules [1, 33]. The distinction is important here because the Recommendation makes an implementation problem visible without enacting the solution proposed in this article.

Paragraph 17 supplies the cross-source premise. It calls for the Recommendation's ethical principles to be applied consistently to neurotechnology and to neural, indirect neural, and non-neural data that allow inferences about mental states. The wording reaches beyond a device-defined approach and permits functional comparison across informational pathways. Consistent principles do not, however, require identical legal treatment for every input or use.

Paragraph 83 addresses differentiation after entry by recommending agile, risk-based, and tiered regulatory frameworks. Any such framework needs variables that distinguish cases once they are covered. Source is relevant where it tracks invasiveness or closed-loop intervention, but it cannot carry the whole classification if paragraph 17's cross-source concern is to matter in practice. Consequential use, secrecy, coercion, vulnerability, reversibility, validity, scale, and institutional power must also shape the response.

Paragraphs 85 and 86 move from scope to data policy. They ask Member States to consider neural, indirect neural, and non-neural data enabling mental-state inferences as sensitive personal data and recommend targeted safeguards. They also address tying access to goods or services, sharing without an appropriate legal basis or explicit opt-in, and targeted advertising without prior free and informed consent [1]. The provisions thereby connect classification to concrete power relations: a refusal right has little value when refusal produces substantial disadvantage or no functionally equivalent alternative exists.

What the Recommendation leaves open is just as important. It specifies no common operational trigger, no separate duties for inputs, outputs, and use, and no three-tier duty structure or presumptive-prohibition procedure. Those institutional choices belong to this article, not to latent rules recovered from UNESCO by interpretation.

Taken together, the provisions support a limited normative bridge. Paragraph 17 weighs against confining mental-state-inference ethics to direct neural measurement; paragraphs 85 and 86 connect that wider scope to sensitive-data safeguards and coercive commercial practices; and paragraph 83 calls for agile, risk-based tiers. They support, but do not enact, two design constraints: coverage should not be defeated by a change of source, and obligations should remain proportionate to the features of the practice. The next task is to define the regulated objects without reifying model outputs or erasing neural-specific risks.

## 5. Three Regulatory Objects and Interactional Provenance

The common trigger applies only when three elements are present. First, technological processing produces or materially shapes a score, label, ranking, vector, profile, alert, adaptive state, or other representation, or is configured prospectively to do so. Second, the representation is presented or operationally treated as evidence about how a person thinks, feels, attends, remembers, intends, decides, exercises volitional control, or is psychologically disposed. Design, configuration, documentation, interface, recipient instruction, and actual deployment can establish this element; the provider's label cannot defeat it. Third, the representation is linked or reasonably linkable to an identified or identifiable person. For a configured process that has not yet produced an output, first-object duties attach only where elicitation or processing is directed to producing a representation satisfying the second and third elements. General technical capability is insufficient.

A closely related mental characteristic is limited to a claimed cognitive or psychological capacity, disposition, intention, or vulnerability that purports to describe or predict mental functioning or reflective choice. A behavioural, demographic, physiological, or preference variable does not qualify merely because it correlates with such a characteristic. Conversely, calling an output engagement, risk, propensity, personalisation, or an embedding does not avoid coverage where its documented or actual function assigns it the covered mental meaning. The trigger requires neither proof that the attribution is true nor a harmful downstream use. Coverage confers neither validity nor admissibility.

Three stage events determine when particular duties attach. Designing or processing inputs and elicitation before an output exists creates preventive first-object duties. Generating a person-linked attribution completes the output trigger and adds second-object duties; retention or transmission renews them. Using the attribution for a decision, intervention, or personalisation adds third-object duties. This division avoids defining the trigger circularly by reference to objects that have not yet been specified.

The first regulatory object is **input and elicitation conditions**. Inputs include neural signals, facial or vocal features, smartphone behaviour, interaction traces, written language, and other data configured for mental-state inference. Elicitation conditions include the task, prompt, stimulus, interface, training procedure, calibration sequence, and incentives through which the input is produced. Duties can arise before any attribution is retained. The proposed rule would therefore apply if a system were configured to restructure an interaction covertly in order to elicit diagnostic or personality-relevant responses. This is a regulatory hypothetical, not a claim that the cited large language model (LLM) study tested such an end-to-end capability.

Inputs are not neutral raw material, and the conditions of elicitation are part of what an output can mean. Han and Chen's conceptual reconstruction shows how a BCI classification may depend on an experimental task and person-specific calibration [31, pp. 6–11]. In Peters and Matz, estimates were generated from up to 200 archived Facebook status updates using fixed prompts and two model versions; accuracy varied by trait, message volume, age, and gender [19, Methods and Results, Fig. 2]. Barrett and colleagues likewise show that facial movement–emotion mappings vary across contexts, persons, and cultures [16, pp. 1–2]. Regulating only the final output would conceal the conditions under which its alleged mental meaning was produced.

The second object is the **mental-state attribution output**. It includes a score, label, ranking, vector, profile, alert, or other representation that can be linked to a person and is presented as bearing on a mental characteristic. The output may be probabilistic and need not be intelligible to the affected person. Retention, transmission, sale, combination, and incorporation into a model or dossier can each activate second-object duties. A transient output can also qualify if it guides an immediate action.

Attribution-specific coverage has an immediate legal consequence: it confers neither evidentiary validity nor admissibility. A person can challenge an inference of depression, hostility, attention, political disposition, or intent without conceding that the output captured a fact lodged inside the mind. Indeed, inadequate construct definition, validation, or provenance may be the reason to exclude a covered attribution from use. Inaccuracy does not defeat coverage when the attribution can still affect the person.

The third object is **system deployment and institutional use**. It covers the use of inputs or outputs for personalisation, recommendation, treatment, discipline, eligibility, pricing, employment, education, insurance, credit, policing, migration, or comparable decisions. Unlike the second object's retention or transmission of a particular output, this object concerns the deployment architecture that organises storage, access, recipients, decision rules, and feedback. Immediate adaptation belongs here even if no stable output is retained. As a regulatory hypothetical, consider a conversational system that altered its persuasive strategy in response to an inferred vulnerability. The cited studies do not establish this end-to-end capability.

Duties follow rather than replace one another across the three objects. Preventive duties survive output generation, second-object duties travel with a retained or transmitted attribution, and third-object duties govern those who organise or act on its use. Compliance at a later stage cannot cure an earlier defect. Downstream consent, for example, does not retrospectively justify covert elicitation, just as a lawful input source cannot validate an unreliable attribution or an impermissible use.

Across all three objects, regulated actors should maintain **interactional provenance**. This article uses that term for records of task design, calibration, participant cooperation, observable resistance, intended context, and material changes in deployment. The term is an author-defined governance concept. It is informed by Han and Chen's conceptual reconstruction of task-dependent and interactive inference [31, pp. 6–11], but it is not borrowed from that account as an established legal category.

The practical value of interactional provenance lies in preserving the conditions that generated an output for validity assessment, context-portability analysis, and contestation. A calibration conducted in a clinical setting should not be presumed fit for employment screening; nor should a platform-derived personality estimate be treated as stable across settings without evidence. Reviewers need to know the task, calibration, intended domain, and performance under non-cooperation or resistance, not merely the confidence score.

Borderline cases should be resolved by applying these elements, not by expanding "mental state" without limit. Outside the trigger are next-word or next-token prediction without person-linked mental meaning; ordinary personalisation based only on expressed choices or observed conduct; aggregate research outputs that are neither reasonably linkable nor used for individual action; unaided human interpretation; and raw physiological measurement without a mental attribution. A practice enters only when technological processing converts it into, or it is operationally used as, a covered person-linked attribution. Thus, a spelling assistant remains outside unless interaction history becomes a depression or personality profile, and a heart-rate report remains outside unless it becomes an attribution of anxiety, attention, or mood. Source does not determine entry; invasiveness, embodiment, monitoring, and closed-loop capacity affect later classification and duties.

For the presumptive-prohibition rule, a **core attribute** is one whose institutional use can substantially affect mental privacy, freedom of thought, identity, or equal status, such as mental health, religious or political commitment, intention, or an identity-forming, highly sensitive, or vulnerability-revealing personality attribution. The term does not cover every personality estimate, preference, or momentary reaction. **Personalised manipulation** means adaptation designed to exploit an inferred susceptibility in order to impair reflective choice; it is narrower than ordinary personalisation or recommendation. These functional definitions control escalation without asserting that the attribution is true.

## 6. Dual-Harm Model, Three Duty Tiers, and Presumptively Prohibited Practices

Entry alone says little about regulatory intensity. That question begins with two harm pathways. The **error pathway** concerns institutional reliance on an invalid, poorly generalised, or context-insensitive inference. Barrett and colleagues identify limits in facial movement–emotion inference [16, pp. 1–2], while Poldrack analyses the logic of neuroimaging reverse inference [20, pp. 59–63]. Neither study establishes that a given error will produce discrimination, stigma, or inappropriate intervention; that consequence depends on the downstream setting. In this model, institutional reliance is the mechanism that connects epistemic weakness to harm.

The **success pathway** concerns the opposite possibility: an attribution may be accurate or useful enough to expose a secret, identify a vulnerability, enable coercion, or support manipulation. Passive smartphone features carried predictive information about future depression-severity scores within Balliu and colleagues' study cohort [18, pp. 6–7], and Peters and Matz found modest trait-score associations in social-media text [19, Results, Fig. 2]. Martinez-Martin and colleagues explain why even ostensibly content-free digital-phenotyping inputs raise privacy and accountability questions [17, pp. 1–3]. Greater predictive usefulness may therefore reduce one kind of error while making an attribution more valuable for surveillance or influence. The cited studies did not test those downstream uses.

Neither pathway determines the applicable duties on its own. Both are assessed against two series that may apply together. **Epistemic duties** concern validity, uncertainty, task dependence, individual calibration, context portability, performance under resistance, contestation, and human review. **Intrusion-and-power duties** concern purpose limitation, restrictions on undisclosed elicitation or use, meaningful refusal, editing or selective disclosure, anti-manipulation safeguards, access controls, and collective oversight. Better validation cannot make a highly accurate covert inference acceptable, and stronger privacy controls cannot rescue a consensual but invalid clinical classifier.

Everything that follows in this section is the article's normative proposal, not a taxonomy found in the UNESCO Recommendation or existing legislation. UNESCO recommends agile and risk-based tiered frameworks but prescribes neither their number nor their thresholds [1, para. 83]. Here, that general direction is translated into three cumulative levels of duty for permitted practices and a separate category of presumptively prohibited practices. Prohibition is kept outside the tiers because it changes the question from which safeguards apply to whether the practice may proceed at all.

### 6.1 Fixed Decision Order

The model uses seven questions in a fixed order. Presumptive prohibition is tested first; a permitted practice then receives the highest applicable duty tier. The core-attribute and personalised-manipulation definitions in Section 5 are additional predicates used within the prohibition test, not hidden substitutes for the seven questions:

1. Does the common trigger apply?
2. Is there a non-consensual closed-loop intervention?
3. Is the elicitation, inference, or institutional use undisclosed or not reasonably apparent to the affected person at the relevant time?
4. Is the attribution used for a major decision affecting access or status in employment, education, insurance, credit, health care, policing, or an equivalent domain?
5. Would refusal predictably cause substantial disadvantage because the actor controls access to a necessary or important benefit and no functionally equivalent alternative exists?
6. Does a heightened-safeguard condition exist through invasive acquisition, close embodied integration, continuous monitoring, third-party exposure of a core attribute, or vulnerability arising from age, health, or institutional dependence?
7. Is a serious and not readily reversible intervention error reasonably foreseeable?

Section 6.5 supplies the materiality and authorisation elements for Q2; iteration alone is insufficient. The questions are branches, not points. Q1 No places a practice outside this model. Q2 Yes yields presumptive prohibition. Q3 records a covert condition, while Q5 records coercive dependence even when the practice is disclosed. Either condition yields presumptive prohibition only when the attribution concerns a core attribute and either Q4 is Yes or personalised manipulation is present. Q5 otherwise yields Tier 3, as does Q4. A Yes answer to Q7 also yields at least Tier 3. Absent one of those higher branches, Q6 yields at least Tier 2; otherwise Tier 1. A Q3-Yes practice that fails the conjunctive prohibition rule remains at the otherwise applicable tier; this neither deems it voluntary nor cures nondisclosure.

These are provisional legal decision rules, not validated risk scores. Table 1 records the branch, evidence, responsible actor, and reclassification condition; sectoral law may require more. Duties remain cumulative across objects and lower tiers. An exception seeker bears the burden of legal authority, necessity, proportionality, and no less restrictive alternative.

**Table 1. Minimum operational record for the seven-question procedure**

| Question | Minimum record | Decision, responsibility, and review |
|---|---|---|
| Q1. Common trigger | Section 5 elements; configuration, assigned meaning, actual use, and person-linkage | Controlling actor; deployer where use assigns meaning. No means outside. Reassess after material change in meaning, linkage, recipient, or use. |
| Q2. Non-consensual closed-loop intervention | Section 6.5 control, materiality, authorisation, and alternatives | Loop deployer or operator. Yes means presumptive prohibition; an external authority decides any ordinary exception and retrospectively reviews any time-limited emergency action. Reassess after changed control, authorisation, exception, or incident. |
| Q3. Covert condition | Notice content, timing, placement, and contextual visibility | Actor controlling disclosure. Record Yes for the conjunctive test. Reassess at each stage; later notice does not cure an earlier covert stage. |
| Q4. Major decision | Decision workflow, attribution weight, effects, and reviewer authority | Institutional user and final decision-maker. Tier 3 and the major-decision limb. Reassess after a material change in role or consequence. |
| Q5. Coercive dependence | Refusal consequences and practical availability of equivalent alternatives | Gatekeeper controlling access. Tier 3 and the coercive-dependence limb. Reassess when alternatives or dependence change. |
| Q6. Heightened safeguard | Acquisition, embodiment, monitoring, recipients, core attribute, and vulnerability | Actor controlling the condition. At least Tier 2 unless a higher branch applies, with condition-specific duties. Reassess when the condition or safeguards change. |

| Question | Minimum record | Decision, responsibility, and review |
|---|---|---|
| Q7. Serious, not readily reversible error | Failure modes, severity, reversibility, detection time, rollback, and incidents | Actors controlling the failure mode. At least Tier 3. Reassess after technical change, new evidence, incident, or near miss. |

Responsibility follows control, reasonably available knowledge, benefit, mitigation capacity, and foreseeable use. It may be shared without being identical, and contract cannot erase it. A general-purpose model provider is not the validator of every downstream construct. Its duties expand, however, when it designs, markets, or supports the covered function, knows of a specific foreseeable deployment, or controls a material mitigation.

**Table 2. Allocation of evidence and responsibility**

| Actor | Controlling role | Core duties and evidence boundary |
|---|---|---|
| General-purpose model provider | Base model, release, documentation, versions, and mitigations | Record model-level conditions and limits; support traceability and feasible mitigation. This does not establish downstream validity or lawfulness. |
| Application or elicitation controller | Construct mapping, prompts, thresholds, sensing, interaction, and calibration | Define the construct and context; document uncertainty and interactional provenance; provide required notice and input minimisation. |
| Institutional deployer or attribution holder | Purpose, population, recipients, consequences, retention, and transfer | Conduct use-specific validation; classify Q3–Q7; provide review and contestation; limit access, retention, transfer, and reuse. |
| Closed-loop operator or maintainer | Actuation, settings, maintenance, fail-safe behaviour, and emergency response | Control relevant failures; maintain rollback, shutdown, incident reporting, and suspension. |
| Final decision-maker or external authority | Consequential decision or power to verify, authorise, and suspend | Review uncertainty and provenance with competence to depart. External review does not cure missing evidence or transfer primary responsibility. |

Where roles overlap, duties accumulate. Where control is divided, the actors must record who supplies each object-specific item of evidence and safeguard. Model-level evidence and use-specific validation answer different questions; neither substitutes for the other.

### 6.2 Tier 1: Residual Baseline and Voluntary Local Use

Tier 1 is the residual duty baseline where no higher branch applies; its paradigm case is voluntary, person-directed, locally contained use without a major decision, third-party exposure of a core attribute, or serious intervention risk. A user may employ an application to reflect on mood, attention, or habits where processing is local, retention is brief, and outputs are not transferred.

The duty holder is the provider or deployer controlling the relevant object. Epistemic duties require a plain account of the construct, intended context, material uncertainty, and limits of generalisation. Intrusion-and-power duties require notice, purpose limitation, accessible deletion, and a default against third-party transfer. The person should be able to stop the inference without losing an unrelated service. Review may remain internal if the system stays local and no substantial external effect is possible.

Tier 1 still subjects consumer-facing wellness profiling to these duties; any personal-processing exception should require genuinely local control and no institutional access.

### 6.3 Tier 2: Heightened Safeguard Conditions

Tier 2 applies as a minimum where Question 6 identifies invasive acquisition, close embodied integration, continuous monitoring, third-party exposure of a core attribute, or vulnerability arising from age, health status, or institutional dependence, but no Tier 3 or presumptive-prohibition branch applies. Examples include ongoing mood inference in voluntary health support, adaptive assistance for a person who depends on a device, or consensual transfer of a core mental-health attribution outside major decision-making. If attention monitoring affects discipline or educational access, Question 4 instead produces Tier 3. If a serious and not readily reversible intervention error is reasonably foreseeable,

Question 7 also produces Tier 3. Persistent observation, bodily coupling, disclosure, and vulnerability can alter the freedom to refuse or magnify the cost of error.

Epistemic duties include construct and criterion validity appropriate to the stated use, subgroup performance, task dependence, person-specific calibration where relevant, context-portability testing, and performance under non-cooperation or resistance. Material uncertainty must reach the human decision-maker, not only appear in technical documentation. The affected person needs a route to correction, second review, or suspension of automated adaptation.

Intrusion-and-power duties include explicit and renewable permission, granular control over sensing and inference, limits on retention and secondary use, role-based access, and an accessible method to pause or reduce participation. Invasive acquisition adds necessity, minimisation, bodily-safety, and removal requirements. Close embodied integration adds continuity-of-function, user-control, maintenance, and exit safeguards. Third-party exposure of a core attribute requires a specific transfer purpose and recipient restriction. Interactional provenance must record changes in tasks, prompts, calibration, and deployment context. Independent ethics or safety review is appropriate where the intervention can affect health, communication, or bodily functioning. An exception to ordinary consent requirements would require a specific legal basis and evidence that a less intrusive means cannot meet the protected purpose.

### 6.4 Tier 3: High-Impact Gatekeeping and Serious Intervention Risk

Three circumstances move a permitted practice into Tier 3. One is the use of an attribution in a major decision concerning employment, education, insurance, credit, health care, policing, or comparable access and status. Another is coercive dependence: refusal would cause substantial disadvantage and no functionally equivalent alternative exists. The third is a reasonably foreseeable intervention error that would be serious and not readily reversible. Neural or non-neural origin is not the decisive fact. What matters is an actor's power to translate a contested mental attribution into an opportunity, burden, or legal position, or an intervention's capacity to impose a comparably serious consequence before ordinary correction is effective.

All Tier 3 practices require validation for the actual use, independent review, documented thresholds, testing for material subgroup and resistance effects, a route to contestation or suspension, and allocation of responsibility to the actor controlling each stage. A model or device provider retains duties for evidence, documentation, foreseeable misuse, and material changes within its control.

For the gatekeeping and refusal-disadvantage routes, the deploying institution is the primary duty holder because it selects the purpose and acts on the attribution. It must disclose the attribution's role, provide human review with authority to depart from it, justify reliance after a material error or provenance mismatch is identified, and provide a functionally equivalent route without mental-state inference where feasible. Lawful and specific purpose, separation from unrelated records, and prohibition of incompatible reuse also apply. Collective oversight is necessary where a practice affects a workforce, school, insured group, or policed population. A works council, professional body, regulator, court, or other competent institution must be able to examine patterns that individual complaints cannot reveal.

For the serious-intervention route, duties follow technical and operational control rather than an assumed institutional decision-maker. The provider, clinical or other professional deployer where one exists, operator, and maintainer must allocate responsibility for hazard analysis, fail-safe behaviour, rollback or safe shutdown, emergency stopping, independent safety validation, incident reporting, and post-incident suspension. The affected person must have an accessible means to pause the intervention where technically and clinically feasible. Closed-loop capacity without non-consensual operation does not by itself trigger prohibition, but it activates these intervention-specific duties; Q2 governs non-consensual operation and Q7 governs foreseeable serious and not readily reversible error.

Some Tier 3 uses may already be prohibited or restricted by sectoral law. The model does not weaken those rules. It supplies a floor for identifying equivalent gatekeeping or serious intervention risk when existing categories are source-specific or when a general automated-decision rule does not apply because a human formally remains in the loop.

### 6.5 Presumptively Prohibited Practices

Two routes lead to presumptive prohibition. The first is a non-consensual closed-loop intervention. For Q2, a closed loop must generate or update a covered attribution from the person's signals, behaviour, or interaction and use that attribution to select or materially modulate an intervention without a substantively independent human decision interrupting the control path. The intervention must act on bodily functioning, clinical treatment, effective communication, or the decisional environment for a consequential choice, and it must be capable of materially constraining, overriding, inducing, or redirecting bodily action, treatment, communication, or reflective choice. Repetition or adaptation alone is insufficient. This governance concept is distinct from Han and Chen's theoretical account of open and closed sensorimotor loops [31, pp. 8–11].

Q2 is Yes when those elements are present without valid, specific, and revocable authorisation. Supported decision-making counts only where it enables a contemporaneous expression of the person's will and preferences. Surrogate authorisation requires legal authority, incapacity for the relevant choice, a purpose-limited and least-restrictive intervention, and independent review of serious effects. An emergency intervention without prior authorisation remains within Q2 and may proceed, if at all, only under the exception rule below. If immediate action precludes prior independent authorisation, any provisional intervention must be emergency-confined, documented, time-limited, and promptly reviewed by an independent authority empowered to suspend or terminate it. A No answer creates no safe harbour: Q6, Q7, both duty series, and stronger law still apply.

Iteration alone does not bring ordinary interface adaptation, command-responsive accessibility support, generic safety alerts, or low-impact changes to layout, pacing, ordering, wording, or content within Q2. The threshold is crossed only when a covered attribution controls an intervention that meets the materiality test, such as altered stimulation or treatment, suppression or substitution of communication, or manipulation of a consequential choice capable of impairing reflective control. The second route to prohibition combines a Q3 covert condition or Q5 coercive dependence with a core attribute and either a major decision or personalised manipulation. Covert political or religious profiling used for employment exclusion and secret vulnerability profiling used for coercive persuasion are limit cases, not claims about a current LLM system.

The default duty is prohibition. A narrow exception must identify the competent legal authority, protected purpose, necessity, proportionality, independent authorisation, monitoring, time limit, and lack of a less restrictive alternative. The exception seeker bears the evidential burden. Even where an exception is legally available, epistemic duties require the strongest feasible validation and continuous external review. Intrusion-and-power duties require minimisation, strict access controls, auditable logs, collective remedies, and suspension authority outside the deploying organisation.

### 6.6 The Allocation Logic

Allocation begins with Q1, which controls entry. Q2 and the conjunctive rule then govern presumptive prohibition; Q4, Q5, and Q7 govern Tier 3; and Q6 sets a Tier 2 minimum when no higher branch applies. Tables 1 and 2 make this sequence auditable without turning it into a score. Table 3 applies the resulting rules to the matched cases. Whatever the final classification, both duty series and all object-specific duties travel with it. Neural origin is not itself a tier, although source-linked features may set a higher minimum or add duties within the final tier.

**Table 3. Matched-case allocation under the proposed model**

| Case | Operative answers | Final classification | Source-related duty consequence |
|---|---|---|---|
| Pair A, neural score used in employment | Q1 Yes; Q4 Yes | Tier 3 | Gatekeeping duties apply; ordinary non-invasive acquisition adds no further duty merely because the source is neural |
| Pair A, text-derived score used in employment | Q1 Yes; Q4 Yes | Tier 3 | The same gatekeeping duties apply despite the non-neural source |
| B1, neural case plus invasive acquisition | Q4 and Q6 Yes | Tier 3 | Tier 3 remains the highest class; necessity, minimisation, bodily-safety, and removal duties are added |
| B2, neural case plus close embodied integration | Q4 and Q6 Yes | Tier 3 | Tier 3 remains the highest class; continuity-of-function, user-control, maintenance, and exit duties are added |
| B3, neural case plus closed-loop capacity | Q4 Yes; Q7 depends on foreseeable error; Q2 depends on non-consensual operation | Tier 3 in the matched employment case; presumptively prohibited if Q2 becomes Yes | Fail-safe, rollback, emergency-stop, independent safety-validation, incident-reporting, and suspension duties are added; capacity alone is not equated with non-consensual intervention |

The matched employment case creates a ceiling effect: Pair A is already Tier 3, so every variant remains at least there. B1–B3 therefore demonstrate stronger duties within the tier rather than a fictitious move above it. From a lower-impact baseline, invasive acquisition or close embodied integration would instead set a Tier 2 minimum through Q6, while serious intervention error would set a Tier 3 minimum through Q7.

## 7. Functional Typology Stress Test

The stress test compares regulatory functions rather than the overall quality of legal systems. It asks where each instrument locates coverage and what remains for interpretation or prospective reform. Their different legal status must remain visible: the UNESCO Recommendation is non-binding; the General Data Protection Regulation (GDPR) and the European Union (EU) AI Act are binding within their respective scopes; the California and Colorado provisions are state statutes; and the Chilean rule is constitutional.

### 7.1 Cross-Source Guidance and General Data Protection

UNESCO reaches furthest across sources. Paragraph 17 places neural, indirect neural, and non-neural data enabling mental-state inference within a consistent ethical frame [1]. Paragraph 83 recommends agile and risk-based tiered frameworks, while paragraphs 85–86 connect that breadth to sensitive-data safeguards. The Recommendation supplies normative scope but little institutional specification: it assigns neither duty holders and remedies nor the three tiers and prohibition rules proposed here.

The GDPR works without a dedicated mental-attribution category. Its general definition of personal data and its rules on profiling, transparency, purpose limitation, data protection impact assessment, and automated decisions can apply across technologies [4, 14]. Inferences about health, religion, political opinion, or other enumerated matters may receive special-category protection. Processing liable indirectly to reveal an enumerated category may also fall within Article 9, but that does not create a general category for mental-state inference [34, 35]. Article 22, for its part, is not a general prohibition on profiling; it concerns solely automated decisions producing legal or similarly significant effects, subject to exceptions and safeguards [4, 36, 37]. Coverage therefore depends on attributed content, processing arrangement, and decisional effect rather than a unified mental-attribution trigger.

In Pair A, both the neural and text-derived instability scores may count as personal data and profiling. Whether either receives special-category treatment depends on what the score reveals; whether Article 22 applies depends on how the employer structures the decision. Under guidance endorsed by the European Data Protection Board (EDPB), token human review does not by itself remove a decision from Article 22, because meaningful review requires authority, competence, and consideration of relevant data [37]. The proposed Tier 3 adds decision-context validation and contestation regardless of the Article 22 outcome, without displacing the GDPR analysis.

### 7.2 Use-Specific AI Regulation

The EU AI Act regulates through system definitions, prohibited practices, high-risk uses, and specified techniques [5]. Article 3(39) defines an emotion-recognition system by reference to biometric data. Article 5(1)(f) prohibits emotion inference in workplaces and educational institutions subject to medical and safety exceptions, but does not repeat that biometric qualifier. The Commission's non-binding Guidelines interpret the prohibition consistently with Article 3(39) and place written-text content or sentiment analysis outside it; authoritative interpretation remains with the Court of Justice of the European Union [38]. This approach directly targets specified consequential uses and reflects concern about the scientific basis of emotion inference.

On the Commission's current interpretation, comparable mental gatekeeping can enter the Act through different routes depending on whether the system uses biometric indicators or language. The resulting gap is functional, not absolute: other provisions may regulate an LLM or decision system according to purpose, deployment, and risk classification. The hybrid proposed here retains use-based prohibitions while adding a prospective source-neutral trigger for attribution-specific duties.

Article 5 and the definition in Article 3(39) have applied since 2 February 2025. Regulation (EU) 2024/1689 retained 2 August 2026 as the general application date for provisions not otherwise staged. The Council gave final approval on 29 June 2026 to a Digital Omnibus amendment setting deferred application dates of 2 December 2027 for stand-alone high-risk-system obligations and 2 August 2028 for high-risk systems embedded in regulated products [39]. The final act was signed on 8 July 2026 and published as Regulation (EU) 2026/1744 in the *Official Journal* on 24 July 2026 [40, 41]. Under Article 4, it entered into force on 27 July 2026, the third day following publication [40].

### 7.3 Neural-Data Statutes

California's SB-1223 adds neural data to sensitive personal information under the California Consumer Privacy Act and defines it through nervous-system measurement. The text expressly excludes information inferred from non-neural information from the neural-data definition [6]. That exclusion creates a clear source boundary. It does not remove non-neural inferences from all consumer-privacy law, but it prevents them from receiving protection solely by virtue of the neural-data amendment.

Colorado's HB24-1058 adds biological data to the Colorado Privacy Act's sensitive-data definition [7]. Biological data expressly includes neural data but is limited to data used or intended to be used for identification purposes. Neural data is separately defined through nervous-system measurement and device processability. The statute's placement within a broader privacy framework supplies consumer rights and controller duties, while its definitions remain source-proximate. The design treats qualifying neural data as sensitive independently of a formal diagnosis, subject to the statute's other definitional and scope requirements.

These statutes do not purport to resolve Pair A across sources. A qualifying nervous-system measurement may enter the added sensitive-data category, but whether a derived score does so depends on the definitions and scope of each statute; the text-derived score does not qualify merely under the neural-data amendment. The B variants retain source attention for a different reason. B1 adds invasive acquisition, B2 embodied integration, and B3 closed-loop capacity, one at a time. Because employment use has already placed the matched case in Tier 3, the variants activate condition-specific duties within that tier. They show why stronger duties may be justified without suggesting that either statute already implements this model.

### 7.4 Constitutional Protection of Brain Activity

Chile's Law 21.383 amended Article 19(1) to require scientific and technological development to respect life and physical and psychological integrity, and directs legislation governing its use in persons especially to safeguard brain activity and information originating from it [8]. The provision has a different function from consumer privacy or AI regulation. It establishes a constitutional orientation and a source-focused protected domain. It does not itself constitute a general code for non-neural mental-state inference or provide the operational tiers described in this article.

Legal form determines what each instrument can do. Constitutional protection may anchor review and future legislation without specifying controller notices or validation tests; a data statute may confer access and deletion rights while leaving coercive public-sector use unresolved; an AI rule may prohibit a practice without classifying every related output; and a non-binding recommendation may coordinate concepts without creating remedies.

### 7.5 Comparative Result

### Table 4. Functional comparison of selected regulatory types

| Regulatory type | Trigger and legal status | Stress-test result | Gap exposed |
|---|---|---|---|
| UNESCO Recommendation | Non-binding cross-source standard for data enabling mental-state inference and risk-based implementation | Pair A: both sources enter its ethical scope. Pair B: differentiation is recommended but not specified. | No operational trigger, duty allocation, or remedy |
| GDPR | Binding rules for personal data, special categories, profiling, and specified automated decisions | Pair A: both scores may be personal data; further classification depends on content and processing. Pair B: aggravators matter through lawfulness, risk, impact assessment, and sectoral law. | No separate mental-attribution category; effect and automation thresholds vary |
| EU AI Act | Binding prohibited-practice, high-risk-use, and biometric emotion-recognition rules | Pair A: entry can differ under the Commission's current interpretation. Pair B3 matters only when activated in a covered function. | Non-biometric personality or psychological inference may enter through other routes |
| California and Colorado statutes | Binding state privacy rules tied to qualifying nervous-system measurement | Pair A: the neural measurement has a source-specific route; the text score does not qualify under the neural amendment. Pair B: the aggravators explain why source-specific safeguards may remain warranted. | Functionally comparable non-neural inference is outside the added neural category |
| Chile Law 21.383 | Constitutional protection of brain activity and information originating from it | Pair A: the neural source receives explicit constitutional attention. Pair B: aggravators may strengthen that concern but are not operational tiers. | No general operational regime for cross-source attribution |

Table 4 is a map of recurrent gaps, not evidence that the proposed model already exists or would produce better outcomes. General data law lacks a dedicated trigger, neural statutes preserve source boundaries, and use-based AI law may classify adjacent attributions differently. The hybrid responds by joining a narrow attribution trigger to activity- and use-based tiers.

Implementation can occur through sensitive-data interpretation, horizontal legislation, sectoral duties, or regulatory classification. A prohibition rule requires clear legal authority; Tier 2–3 duties can follow existing institutional competence. Allocation for Japan is a separate inquiry.

## 8. Objections and Replies

### 8.1 The Trigger Is Overbroad and Reifies Mental States

Overbreadth is troubling here for a reason deeper than the volume of data covered. Once law names mental-state inference as a category, institutions may treat uncertain outputs as evidence of a stable inner object. The category might also spill into ordinary recommendation, communication, or contextual interpretation. Regulation could then lend pseudoscientific constructs institutional legitimacy merely by naming them.

That risk is why the trigger attaches to neither every prediction nor every human interpretation or reference to mood. Coverage requires technological mediation, an operational function that treats a representation as evidence about a defined mental characteristic, and person-linkage. The test excludes ordinary prediction but prevents evasion through labels such as engagement, propensity, or risk. Barrett and colleagues identify dependencies in facial inference [16, pp. 1–2], Poldrack does the same for reverse inference [20, pp. 59–61, Table 1], and Han and Chen frame BCI decoding as interactional [31, pp. 6–11]. The corresponding institutional safeguard is interactional provenance. Because coverage confers neither validity nor admissibility, a court, regulator, employer, clinician, or platform may have to reject an attribution whose construct, validation, or provenance is inadequate. The trigger imposes a duty to justify or exclude; it is not a licence to believe.

### 8.2 Neural Data Are Distinctive

The neuro-specificity objection holds that neural data implicate the body, agency, and mental integrity in ways that text or behavioural traces do not. A source-neutral category may dilute political attention and legal protection just as invasive or closed-loop systems become practical. Neurorights and brain-data governance emerged for reasons that a general profiling framework cannot absorb [11–13, 21, 24].

Source-neutral entry need not deny neural distinctiveness. Pair B isolates invasiveness, embodiment, and closed-loop capacity in variants B1, B2, and B3. A non-consensual closed-loop intervention that meets Section 6.5's materiality test enters the presumptive-prohibition category without further balancing. Invasive acquisition and close embodied integration set a Tier 2 minimum under Q6 and add condition-specific duties even when another branch has already produced Tier 3. Closed-loop capacity adds intervention-safety duties; materially non-consensual operation activates Q2, while foreseeable serious and not readily reversible error activates Q7. A text-derived personality score used for employment therefore cannot escape attribution-specific scrutiny, but neither is it treated as equivalent to an implanted intervention.

### 8.3 Existing Law Is Sufficient

The redundancy objection points to privacy, data protection, equality, consumer protection, sectoral safety law, and due process. Creating another classification can increase compliance complexity and rights inflation. GDPR scholarship in particular shows that existing principles can address many neurotechnology concerns [14]. A regulatory model should not assume that every conceptual gap requires a new right.

The legal-form premise is sound: implementation can use existing authorities, and a stronger sectoral rule should not be duplicated. The additional work is classificatory and procedural. Existing regimes often activate through proxies such as signal source, enumerated sensitive content, sole automation, a specified biometric technique, or a completed discriminatory effect. The trigger asks a common set of attribution-specific questions before those routes diverge and places epistemic duties alongside intrusion-and-power duties. Privacy compliance may control collection while leaving an invalid construct operational; model validation may improve accuracy while leaving coercive gatekeeping intact.

Redundancy should therefore be assessed duty by duty. If existing law already supplies notice, access, an effective challenge, independent audit, and a prohibition for the same practice, the model adds no parallel procedure. If a human-in-the-loop arrangement avoids an automated-decision rule while the human merely ratifies a psychological score, Tier 3 requires substantive review. If consumer consent is formally obtained through an unavoidable service, Question 5 tests whether refusal was meaningful.

### 8.4 Regulation Should Target Activities and Uses Only

The activity-only objection offers the most serious alternative. It regulates covert manipulation, gatekeeping, and non-consensual intervention without constructing a sensitive data category, and it remains workable when a system infers and acts without retaining an output. Han and Chen's interactional account of BCI-mediated mind reading sharpens this challenge [31, pp. 6–11]. Activity-based regulation should prevail wherever the harmful act can be defined directly and the responsible actor can be assigned the necessary evidence, review, and remedy duties. Intervention, manipulation, and consequential reliance remain principally intervention, use, and decision questions; an equivalent sectoral rule satisfies the corresponding third-object duty.

What remains distinctive is found upstream of the harmful act. At the first object, notice, permission, minimisation, and covert-elicitation limits govern configurations directed to covered attributions. Interactional provenance preserves task, calibration, cooperation, resistance, and intended context before a deployer receives a score. At the second, access, correction, retention, transfer, recipient restrictions, provenance, and purpose limitation govern attributions that can circulate or be repurposed before consequential use. First-object duties require a configuration directed to covered attribution; second-object duties require generation, retention, or transmission. The cited LLM evidence does not establish cross-context retention or deployment.

This division also prevents validation from being duplicated. Upstream actors supply evidence about the construct, generation conditions, and model-level limits within their control; deployers remain responsible for population, context, threshold, decisional role, consequences, and alternatives. The trigger's narrow additionality lies in closing provenance, transfer, and pre-use contestation gaps, while prohibition and high-impact escalation remain activity- and power-based.

**8.5 The Tiers Are Arbitrary and Administratively Unworkable**

The architecture itself invites a final objection. Three duty tiers, a separate presumptive-prohibition category, seven questions, and two escalation predicates may create a false appearance of precision. Terms such as substantial disadvantage, core attribute, personalised manipulation, or serious irreversibility require judgment, and regulated actors may classify strategically. Without empirical calibration, the procedure could burden low-risk practices while missing novel forms of coercion.

That objection is correct about the status of the thresholds: they are provisional legal design, not validated measurements. The three tiers correspond to baseline duties, heightened safeguards, and independent control of high-impact gatekeeping or serious intervention risk. Presumptive prohibition is separate because it addresses permissibility before the level of compliance is considered. The seven questions mark proposed minimum branch points, with the core-attribute and personalised-manipulation predicates completing the conjunctive prohibition rule.

Section 6.5 treats materially bounded non-consensual closed-loop intervention as a presumptive-prohibition bright line. Major decision-making, coercive dependence, and serious, not readily reversible intervention error lead to Tier 3, while Q6 conditions require sector-specific guidance, examples, and periodic evaluation. Implementation evidence might eventually support two or four duty tiers instead. Tables 1 and 2 specify a minimum record and division of responsibility, not empirical calibration. Administrability still depends on sector-specific evidence standards, competent review, and reclassification when provenance or use changes. A jurisdiction unable to provide that review should not authorise a higher-risk practice merely because a decision tree exists.

The hybrid is defensible only while it remains narrow, preserves neural-specific aggravators, defers to stronger law, and reaches harmful activity even when no output is retained. Its thresholds remain open to empirical and participatory calibration.

## 9. Conclusion

Mental-state-inference governance need not choose between neural exceptionalism and an unlimited category of mental data. A neural boundary can miss non-neural attributions used for the same institutional purpose; an unlimited category can obscure uncertainty and turn model-dependent representations into apparent facts about a person's mind. The narrower point of entry proposed here is technologically mediated, person-linked mental-state attribution.

That common entry is followed by differentiation. Duties accumulate across input and elicitation conditions, attribution outputs, and deployment or institutional use, while interactional provenance keeps task, calibration, cooperation, resistance, and intended context attached to the output. Error and success remain distinct harm pathways. Epistemic and intrusion-and-power duties are assessed independently because either pathway may engage both. Seven ordered questions and two escalation predicates then allocate permitted practices across three duty tiers. Materially autonomy-affecting non-consensual closed-loop intervention, together with specified combinations of covert or coercive inference, core attributes, major decisions, and manipulation, is placed in a separate presumptive-prohibition category.

The matched cases show how common coverage can coexist with differentiated obligations. Pair A changes only the neural or non-neural source of a functionally comparable gatekeeping attribution. B1, B2, and B3 add invasive acquisition, embodied integration, and closed-loop capacity separately. Employment use has already placed Pair A in Tier 3, so those variants add condition-specific duties rather than an artificial higher tier. They make the proposed weighting explicit without purporting to prove the effects of each feature. Data, AI, consumer, sectoral, and constitutional regimes could implement that logic through different legal forms.

Several questions remain beyond the article's reach. The analysis does not validate the tier thresholds empirically, exhaust the relevant jurisdictions, settle every taxonomy of mental states, establish current LLM capability beyond the cited evidence, or allocate responsibilities within a particular national system. These are substantive limits rather than residual drafting details. Further work must identify jurisdiction-specific legal authority, calibrate thresholds through empirical and participatory research, assess enforcement capacity, and test whether the duties reduce error, intrusion, and institutional abuse without legitimating unreliable inference.

## Funding

This work was supported by JSPS KAKENHI Grant Number JP26K05287.

## Data Availability

This conceptual and legal analysis generated no research datasets. All materials on which the analysis relies are cited in the article.